\documentclass[preprint,preprintnumbers,amsmath,amssymb]{revtex4}
\usepackage{graphicx}
\usepackage{bm}

\usepackage{mhchem}
\usepackage{bm}
\usepackage{amsmath}
\usepackage{amsfonts}
\usepackage{amssymb}

\usepackage{breqn}

\usepackage{balance}

\usepackage{times,mathptmx}

\newcommand{\id}{{\bm 1}}

\newcommand{\be}{\begin{equation}}
\newcommand{\ee}{\end{equation}}
\newcommand{\bea}{\begin{eqnarray}}
\newcommand{\eea}{\end{eqnarray}}

\begin{document}
\title{Direct evidence of plastic events and dynamic heterogeneities in soft-glasses}
\author{R. Benzi$^{1}$, M. Sbragaglia$^{1}$, P. Perlekar$^{2}$, M. Bernaschi$^{3}$ S. Succi$^{3}$, F.Toschi$^{3,4}$ \\
$^{1}$ Department of Physics and  INFN, University of ``Tor Vergata'', Via della Ricerca Scientifica 1, 00133 Rome, Italy\\
$^{2}$ TIFR Centre for Interdisciplinary Sciences, 21 Brundavan Colony, Narsingi, Hyderabad 500075, India\\
$^{3}$ Istituto per le Applicazioni del Calcolo CNR, Viale del Policlinico 137, 00161 Roma, Italy\\
 Department of Physics and Department of Mathematics and Computer Science and J.M. Burgerscentrum, Eindhoven University of Technology, 5600 MB Eindhoven, The Netherlands}

\begin{abstract} 
By using fluid-kinetic simulations of confined and concentrated emulsion droplets, we investigate the nature of space non-homogeneity in soft-glassy dynamics and provide quantitative measurements of the statistical features of plastic events in the proximity of the yield-stress threshold. Above the yield stress, our results show the existence of a finite stress correlation scale, which can be mapped directly onto the {\it cooperativity scale}, recently introduced in the literature to capture non-local effects in the soft-glassy dynamics. In this regime, the emergence of a separate boundary (wall) rheology with higher fluidity than the bulk, is highlighted in terms of near-wall spontaneous segregation of plastic events. Near the yield stress, where the cooperative scale cannot be estimated with sufficient accuracy, the system shows a clear increase of the stress correlation scale, whereas plastic events exhibit intermittent clustering in time, with no preferential spatial location. A quantitative measurement of the space-time correlation associated with the motion of the interface of the droplets is key to spot the long-range {\it amorphous} order at the yield stress threshold.
\end{abstract}

\maketitle

\section{Introduction}
Soft amorphous materials, such as emulsions, foams, microgels and colloidal suspensions, display complex flow properties, intermediate between the solid and the liquid state of matter: they are solid at rest and able to store energy via elastic deformation, whereas they flow whenever the applied stress exceeds a critical yield threshold. The yielding behavior makes such systems as interesting for applications as challenging from the fundamental point of view of out-of-equilibrium statistical mechanics \cite{Larson}. Some of these systems, referred as {\it simple yield stress fluids} (including nonadhesive emulsions and microgels), were shown to flow via a sequence of reversible elastic deformations and local irreversible plastic rearrangements, associated with a microscopic yield stress. These physical ingredients lie at the core of mesoscopic models for soft-glassy dynamics \cite{Sollich1,Sollich2,Fielding1,Fielding2,Hebraud,Picard,Bocquet,Pouliquen,Mansard13}. A challenging question concerns the emergence  of features that are non-homogeneous in space  (like, for example, shear bandings), where the {\it global} rheology is unable to properly capture the complex space-time behavior of the system. One needs to properly bridge between {\it local} and global rheology of the soft-glasses, an issue that has been recently addressed in several papers \cite{Goyon1,VanHecke,Softmatter}. In \cite{Goyon1,Goyon2,Softmatter,Bocquet} it was suggested that such a bridge can be established by introducing a cooperativity scale which determines correlations (non-local effects) in the flow rheology. The underlying idea is that correlations among plastic events exhibit a complex spatio-temporal scenario: they are correlated at the microscopic level with a corresponding cooperativity flow behavior at the macroscopic level. It is the aim of this paper to study the nature of space non-homogeneity in soft-glassy dynamics and to understand the link with correlations emerging from the dynamics of plastic events. More precisely, we investigate the above issues by using a mesoscopic approach based on the Lattice Boltzmann method \cite{CHEM09,EPL10,recentEPL}, which allows the simulation of emulsion droplets and their interface motion under different load conditions. The simulations provide access to a broad spectrum of scales of motion at a very competitive computational cost, a fact that is instrumental for large-scale simulations of yielding materials, where the dynamics of a collection of a substantial number of droplets needs to be accounted for. The peculiar features of plastic events are investigated below and above the yield stress threshold. Above the yield stress, the ``fluidity'' model recently introduced by Goyon {\it et al}.\cite{Goyon1,Goyon2,Jop,Mansard13} captures the essential features of the flow: fluidity changes near the boundaries on a scale $\xi$ which is close to the stress correlation scale and to the characteristic scale of plastic events. Near the yield stress, however, the cooperativity scale can not be estimated with enough accuracy, whereas the stress correlation scale shows a clear increase. In this regime, plastic events do not show any preferential location and the system starts to behave as an elastic medium, characterized by near zero fluidity (i.e. large viscosity) and with a long-range {\it amorphous} order. Our findings echo some recent results on slowly driven thermal glasses\cite{Chikkadi11} and on driven athermal amorphous materials \cite{Maloney,Olson}.

\section{Dynamic rheological model}\label{intro}

We resort to a lattice kinetic model that has already been described in several previous papers \cite{CHEM09,EPL10}. Here, we just recall its basic features. We start from a mesoscopic lattice Boltzmann model for non ideal binary fluids, which combines a small positive surface tension, promoting highly complex interfaces, with a positive disjoining pressure, inhibiting interface coalescence. The mesoscopic kinetic model considers two fluids $A$ and $B$, each described by a {\it discrete} kinetic distribution function $f_{\zeta i}({\bm r},{\bm c}_i;t)$, measuring the probability of finding a particle of fluid $\zeta =A,B$ at position ${\bm r}$ and time $t$, with discrete velocity ${\bm c}_i$, where the index $i$ runs over the nearest and next-to-nearest neighbors of ${\bm r}$ in a regular two-dimensional  lattice \cite{BCS,CHEM09}. In other words, the mesoscale particle represents all molecules contained in a unit cell of the lattice. The distribution functions evolve in time under the effect of free-streaming and local two-body collisions, described, for both fluids ($\zeta=A,B$), by a relaxation towards a local equilibrium ($f_{\zeta i}^{(eq)}$) with a characteristic time scale $\tau_{LB}$:
\begin{dmath}
\label{LB}
f_{\zeta i}({\bm r}+{\bm c}_i,{\bm c}_i;t+1) -f_{\zeta i}({\bm r},{\bm c}_i;t)  = -\frac{1}{\tau_{LB}} \left(f_{\zeta i}-f_{\zeta i}^{(eq)} \right)({\bm r},{\bm c}_i;t)+F_{\zeta i}({\bm r},{\bm c}_i;t).
\end{dmath}
The equilibrium distribution is given by \be f_{\zeta i}^{(eq)}=w_i
\rho_{\zeta} \left[1+\frac{{\bm u} \cdot {\bm c}_i}{c_s^2}+\frac{{\bm
      u}{\bm u}:({\bm c}_i{\bm c}_i-c_s^2 \id)}{2 c_s^4} \right] \ee
with $w_i$ a set of weights known a priori through the choice of the
quadrature \cite{Sbragaglia07,Shan06}. Coarse grained hydrodynamical
densities are defined for both species $\rho_{\zeta }=\sum_i f_{\zeta
  i}$ as well as a global momentum for the whole binary mixture ${\bm
  j}=\rho {\bm u}=\sum_{\zeta , i} f_{\zeta i} {\bm c}_i$, with
$\rho=\sum_{\zeta} \rho_{\zeta}$. The term $F_{\zeta i}({\bm r},{\bm
  c}_i;t)$ is just the $i$-th projection of the total internal force
which includes a variety of interparticle forces. First, a repulsive
($r$) force with strength parameter ${\cal G}_{AB}$ between the two
fluids
\begin{equation}
\label{Phase}
{\bm F}^{(r)}_\zeta ({\bm r})=-{\cal G}_{AB} \rho_{\zeta }({\bm r}) \sum_{i, \zeta ' \neq \zeta } w_i \rho_{\zeta '}({\bm r}+{\bm c}_i){\bm c}_i
\end{equation}
is responsible for phase separation \cite{CHEM09}.  Furthermore, both fluids are also subject to competing interactions whose role is to provide a mechanism for {\it frustration} ($F$) for phase separation \cite{Seul95}. In particular, we model short range (nearest neighbor, NN) self-attraction, controlled by strength parameters ${\cal G}_{AA,1} <0$, ${\cal G}_{BB,1} <0$), and ``long-range'' (next to nearest neighbor, NNN) self-repulsion, governed by strength parameters ${\cal G}_{AA,2} >0$, ${\cal G}_{BB,2} >0$)
\begin{dmath}\label{NNandNNN}
{\bm F}^{(F)}_\zeta ({\bm r})=-{\cal G}_{\zeta \zeta ,1} \psi_{\zeta }({\bm r}) \sum_{i \in NN} w_i \psi_{\zeta }({\bm r}+{\bm c}_i){\bm c}_i -{\cal G}_{\zeta \zeta ,2} \psi_{\zeta }({\bm r}) \sum_{i \in NNN} w_i \psi_{\zeta }({\bm r}+{\bm c}_i){\bm c}_i
\end{dmath}
with $\psi_{\zeta }({\bm r})=\psi_{\zeta }[\rho({\bm r})]$ a suitable pseudo-potential function \cite{SC1,SbragagliaShan11}. Despite their inherent microscopic simplicity, the above dynamic rules are able to promote a host of non-trivial collective effects \cite{CHEM09,EPL10}.  By a proper tuning of the phase separating interactions (\ref{Phase}) and the competing interactions (\ref{NNandNNN}), the model simultaneously achieves small positive surface tension $\Gamma$ and positive disjoining pressure $\Pi_d$. This allows th simulations of droplets of one dispersed phase into the other (see left panel of figure \ref{fig:3}) which are stabilized against coalescence. Once the droplets are stabilized, different packing fractions and polydispersity of the dispersed phase can be achieved. In the numerical simulations presented in this paper, the packing fraction of the dispersed phase in the continuum phase is kept the same and approximately equal to $90 \%$. The model gives direct access to the hydrodynamical variables, i.e., density and velocity fields, as well as the local (in time and space) stress tensor in the system, the latter characterized by both the viscous (fluid) as well as the elastic (solid) contributions. Thus, it is extremely useful to properly characterize the relationship between the droplets dynamics, their plastic rearrangements, and the stress fluctuations \cite{Softmatter}.

\section{Numerical evidence of plastic events}\label{plastic}

\begin{figure}[t!]
\centering
\includegraphics[scale=0.68]{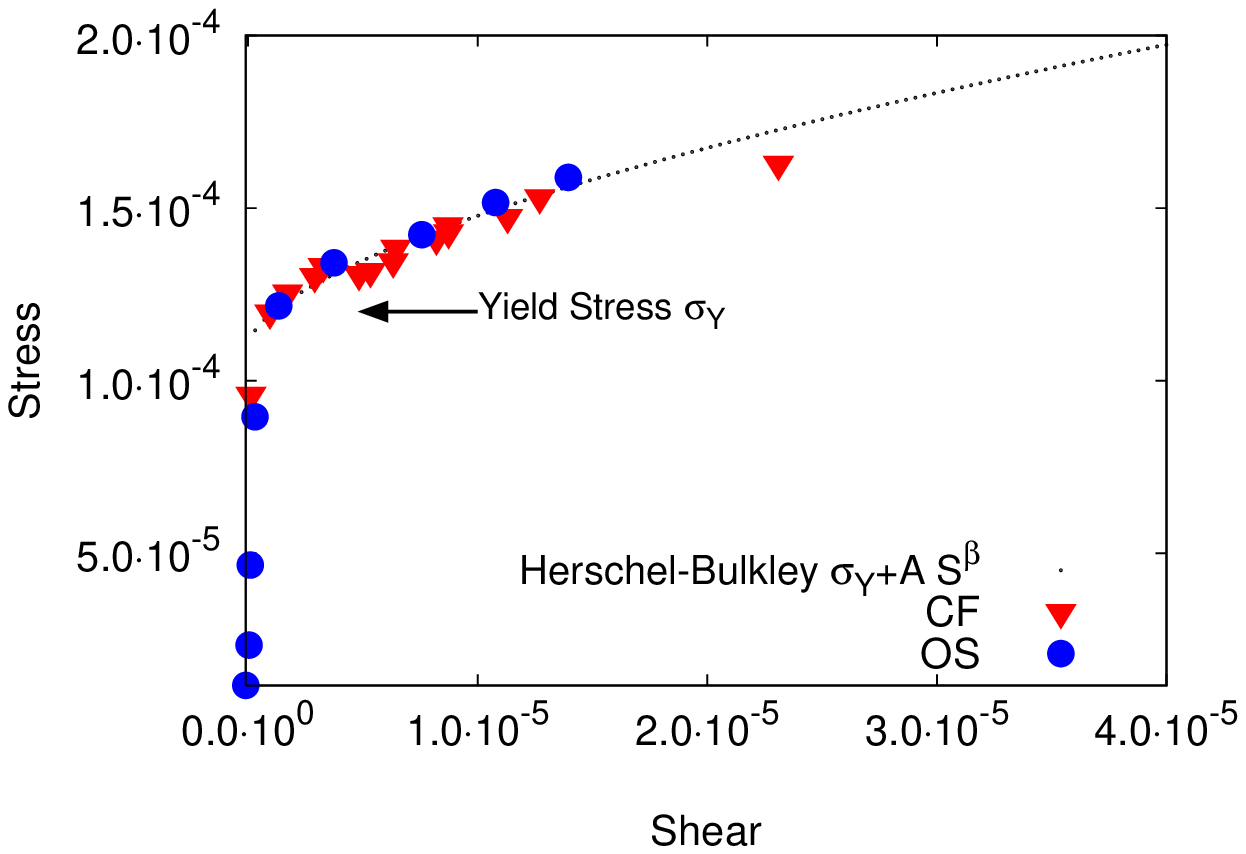}
\includegraphics[scale=0.68]{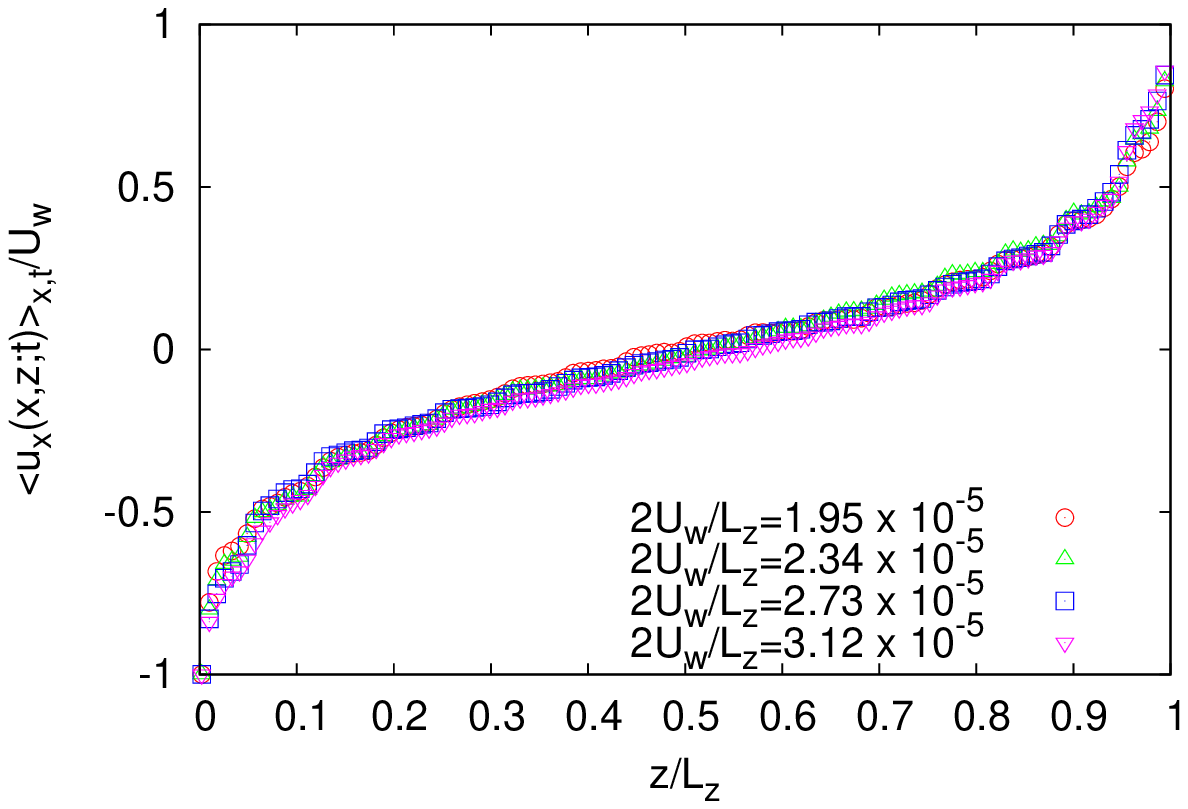}
\hfill
\caption{Top Panel: the plot shows the shear-stress relation for the two flows discussed in the paper. The inverted triangles refer to Couette Flow (CF), whereas the filled circles to the Oscillating Strain (OS). The dotted line represents the Herschel-Bulkley (HB) fit. The analysis in this paper is focused on flows characterized by a stress either slightly smaller or slightly larger than the yield stress. Bottom Panel: normalized velocity profiles, $U(z)/U_W$, as a function of the reduced position $z/L_z$ in a confined steady Couette geometry for different values of the nominal shear $2U_W/L_z$. The packing fraction of the dispersed phase in the continuum phase is kept the same and approximately equal to $90 \%$. \label{fig:1}}
\end{figure}

\begin{figure}[h]
\begin{center}
\centering
\includegraphics[scale=0.5]{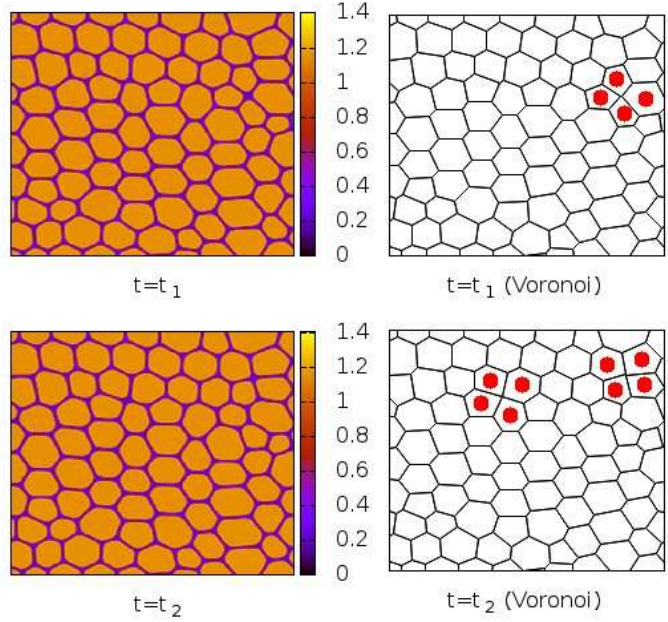}
\caption{Identification of plastic events by using the Voronoi algorithm (see text for details). The analysis reported in this figure is carried out in a time interval $[0:300]$ (in units of $1000$ lbu). Left panel: two time snapshots are reported with blue/yellow (dark/light) colors indicating $A$-rich/$B$-rich regions. Right panel: we report the corresponding Voronoi tessellation of the centers of mass of the droplets. The Voronoi cells involved in the plastic event are labeled by a central dot. The plastic rearrangement at $t_1=263$ is generating a perturbation that affects the successive plastic rearrangements at $t_2=268$. \label{fig:3}}
\end{center}
\end{figure}

To place our results within the proper perspective, we first analyze the global rheological properties of the system under investigation. The computational domain is a rectangular box of size $L_x \times L_z$ ($x$ is the stream-flow direction) covered by $N_x \times N_z =1024 \times 1024$ lattice sites. The simulations, performed on latest generation Graphics Processing Units (GPU) \cite{GPU}, require a few GPU hours for one million time steps, the typical time span of a single run. All quantities will be given in lattice Boltzmann units (lbu) and brackets $\langle ... \rangle$ will be used to indicate averages, either in time ($\langle ... \rangle_{t}$), in space (($\langle ... \rangle_{x,z}$)), or both. Two different boundary conditions are considered: (a) planar Couette Flow with steady velocity at the boundaries $\pm U_W$; (b) Oscillating Strain conditions with strain $\gamma(t) = \gamma_p \sin( \omega t)$ and boundary velocity $U(t) = L_z {\dot \gamma}(t)$. In figure \ref{fig:3} we show a zoom of the configurations resembling the initial conditions (the same for both boundary conditions). For the Oscillatory Strain boundary conditions, the frequency $\omega$ is chosen to guarantee that the stress, $\sigma(t)$, and the strain, $\gamma(t)$, are homogeneous in $z$ for very small $\gamma_p$. We then write $\sigma(t) = \sigma_p \sin( \omega t + \phi)$, where $\sigma_p$ denotes the maximum value of $\sigma(t)$. In figure \ref{fig:1} we show the resulting shear-stress relation following the definition of the global shear $S$ and stress $\sigma$ discussed in \cite{recentEPL}. Our simulations provide a yield stress value of about $\sigma_Y \sim 1.2 \times 10^{-4}$ lbu independently of the two load conditions used. The stress is compatible with a Herschel-Bulkley (HB) relation \cite{Larson}
\begin{equation}
\label{HB}
\sigma  = \sigma_Y + A S^{\beta}
\end{equation}
with $\beta \sim 0.61$. Thus, the material in point shows a non-trivial rheology. The bottom panel of figure \ref{fig:1} reports the normalized velocity profiles, $U(z)/U_W$, as a function of the reduced position $z/L_z$ in a confined steady Couette Flow for different values of the nominal shear $2U_W/L_z$. In absence of non-local effects, one would expect the reduced velocity profiles to be a straight line. This is however not the case: the normalized profiles collapse on the same master curve independent of the applied shear, and emphasize that the shear rate is greater at the wall than in the center of the channel. This non-local effect has been discussed in terms of plastic rearrangements of the flow \cite{Goyon1,Goyon2,Mansard13,Softmatter}. It is therefore of great interest to provide {\it direct} dynamic evidence of such plastic events.

\begin{figure}[h!]
\begin{center}
\includegraphics[width=0.44\textwidth]{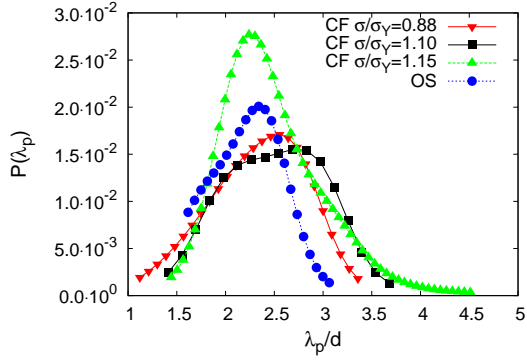}
\caption{Probability density function $P(\lambda_p)$ computed for the plastic events in pre-yield conditions for the Couette Flow (CF) and the Oscillatory Strain (OS) numerical simulations (see text for details). The quantity $\lambda_p$ refers to the characteristic spatial scale of the plastic events and is computed as the square root of the area of the droplets involved in the events (see figure \ref{fig:3}). In both cases $P(\lambda_p)$ is peaked around $\lambda_p \sim (2.0-2.5)d$, where $d$ is the average droplet diameter. \label{fig:4}}.
\end{center}
\end{figure}

To develop a systematic analysis of plastic events, we perform a Voronoi tessellation\footnote{The Voronoi tessellation has been performed by using the Voro++ library. https://math.lbl.gov/voro++} constructed from the centers of mass of the droplets, a representation which is particularly well suited to capture and visualize plastic events in the form of droplets rearrangements and topological changes, occurring within the material.  Such events are shown in the right panel of Figure \ref{fig:3}. The involved Voronoi cells are labeled by a central dot. Quite often, multiple plastic events are observed to take place in short sequence, as evidenced in the bottom-right panel of Figure \ref{fig:3}. Next, we used the Voronoi tessellation to analyze the statistical distribution $P(\lambda_p)$ of the characteristic scale $\lambda_p$ of plastic events below and above the yield stress. Here, $\lambda_p$ is defined as the square root of the area of the droplets involved in the plastic event. In figure \ref{fig:4}, we show $P(\lambda_p)$ as a function of $\lambda_p/d$, where $d$ is the average droplet diameter.  In all cases, $P(\lambda_p)$ shows a well defined peak around $ \lambda_p \sim (2.0-2.5)d$, which corresponds approximately to T1 events involving four droplets. We also note that the tail of $P(\lambda_p)$ gets relatively fatter at large $\lambda_p$ as the average stress is increased, namely for the case $\sigma/\sigma_Y = 1.15$, suggesting that more and more droplets are involved in plastic events.


\begin{figure}[h!]
\begin{center}
\includegraphics[width=0.44\textwidth]{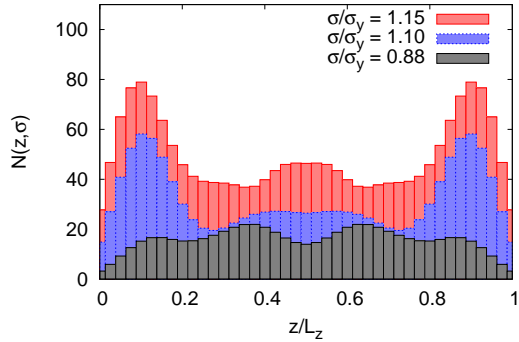}
\caption{Plastic events and their location in a Couette Flow simulation. We report the number of plastic events $N(z,\sigma)$ which occur, for any $x$,  at location $z/L_z \in [0:1]$ for $\sigma/\sigma_Y=0.88,\, 1.1, \, 1.15$. Below the yield stress plastic events are distributed almost uniformly in $z$ whereas for $\sigma>\sigma_Y$ there exists a preferential location near the boundary  with characteristic thickness of the order of $0.2\,L_z$, which is close to $\lambda_p$, i.e., the characteristic scale of plastic events (see text for details). \label{fig:5}}
\end{center}
\end{figure}

Next, we analyze plastic events and their space distribution, to characterize the transition at the yield stress. In particular, we consider the Couette Flow and compute the number of plastic events $N(z,\sigma)$ which occur, for any $x$, at location $z/L_z \in [0:1]$. Results are displayed in figure \ref{fig:5} for different values of the average stress $\sigma/\sigma_Y=0.88,\, 1.1,\, 1.15$. The clear feature emerging from figure \ref{fig:5}, is that below the yield stress ($\sigma/\sigma_Y=0.88$), plastic events are distributed almost uniformly in $z$, whereas for $\sigma>\sigma_Y$ there exists a preferential location near the boundary, with a characteristic thickness of the order of $0.2\,L_z$. It turns out that such thickness is also close to $2\lambda_p$. Thus, two main messages are conveyed by figures \ref{fig:4} and \ref{fig:5}: most plastic events show the same characteristic scale $\lambda_p$, while their number increases by increasing $\sigma$; above the yield stress, a preferential concentration of plastic events occurs near the boundaries in a layer of thickness $2\lambda_p$. Although it is not surprising that most of the plastic events concentrate near the boundaries, the fact that for $\sigma < \sigma_Y$ this does not occur, appears to be non-trivial.


\section{Connection with Fluidity Model}\label{fluidity}

Hereafter, we consider the results of section \ref{plastic} and establish a connection between the plastic events in droplets rearrangements and the corresponding cooperative flow behavior at the hydrodynamic scale. A step towards this goal has been taken in recent works\cite{Goyon1,Goyon2,Bocquet}, where the rate of plastic events is connected to the ``fluidity'' field, defined as the ratio between  the shear rate and the stress,  $f=\frac{\dot{\gamma}}{\sigma}$. By using a kinetic model for the elasto-plastic dynamics of the stress distribution function, the local fluidity is shown to obey (in the steady state) the following equation \be\label{eq:fluidity}
\begin{split}
\xi^2 \Delta f+\left(f_b-f \right)=0
\end{split}
\ee
where the scale $\xi$ is a measure of the non-locality of the cooperativity within the flow. The quantity $f_{b}$ is the bulk fluidity, i.e. the value of the fluidity in the absence of spatial heterogeneities. The bulk fluidity depends upon the local shear rate only, whereas $f$ depends upon the position in space. Its value is equal to $f_b$ when the stress and the shear rate are constant in an unbounded geometry, i.e. without the perturbing effects of the boundaries. The fluidity model has been tested with considerable success both in experiments \cite{Jop,Goyon1,Goyon2} and in molecular dynamics simulations \cite{Mansard13}. Under the hypothesis of low cooperativity, the model predicts proportionality between the fluidity and the rate of plastic events \cite{Jop,Mansard13}. This feature is strikingly robust, as also evidenced by the work of Nicolas \& Barrat, based on a different mesoscopic model of interacting elasto-plastic blocks \cite{Nicolas}. Thus, an increase of the number of plastic events near the boundary should be correlated to a corresponding  increase of the fluidity. Also, we may argue that the cooperativity scale $\xi$ should be of the order of $\lambda_p$, a statement that echoes the results presented by Mansard {\it et al.} \cite{Mansard13}, where molecular dynamics simulations with the ``bubble model'' of Durian\cite{Durian} were compared with the fluidity model. A good agreement was found by using a value of $\xi$ of the order of 5 bubbles radii. To check the validity of this interpretation, we investigate the behavior of the Couette Flow {\it above the yield stress} $\sigma_Y$. In this case, the mean shear stress is spatially homogeneous, which considerably simplifies the solution of equation (\ref{eq:fluidity}).  We first consider the fluidity averaged in time and in the stream-flow direction. For such a $1d$ case, the fluidity is predicted to obey a non-local equation of the form \cite{Goyon1,Goyon2}
\be\label{eq:fluidity1}
\xi^2 \frac{d^2 f(z)}{d z^2}+(f_{b}(\sigma)-f(z))=0
\ee
where $\sigma$ (and hence $f_b(\sigma)$) is a constant in the
stationary Couette flow. The solution of the fluidity equation
requires boundary conditions, i.e. one has to prescribe the value of the fluidity close to the boundaries. When the boundary condition is the same, $f(0)=f(L_z)=f_w$, the expression of the shear rate $\dot{\gamma}=\sigma f$ reduces to:
\be\label{eq:fluidityCF}
\dot{\gamma}(z)=\sigma \left(f_{b}(\sigma)+(f_w-f_{b}(\sigma))\frac{\cosh((z-L_z/2)/\xi)}{\cosh(L_z/2\xi)} \right).
\ee


\begin{figure*}[t]
\includegraphics[scale=0.42]{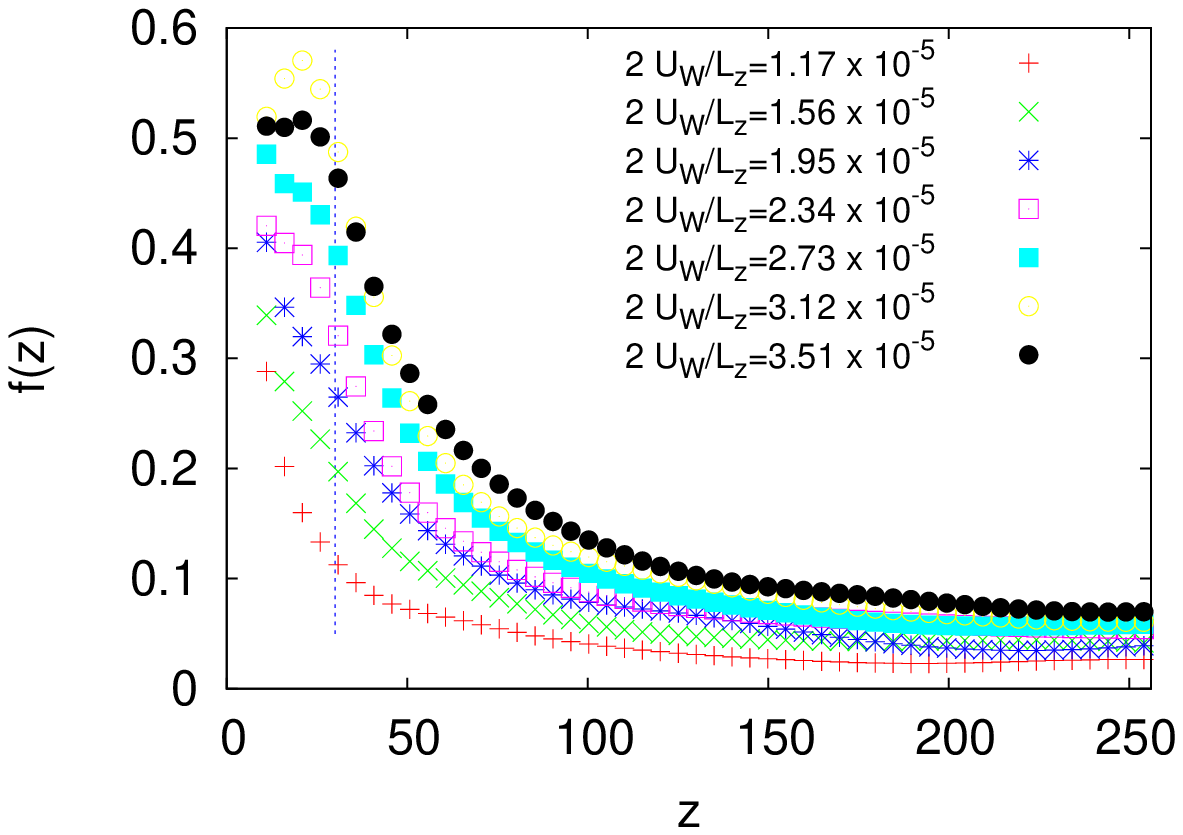}
\includegraphics[scale=0.42]{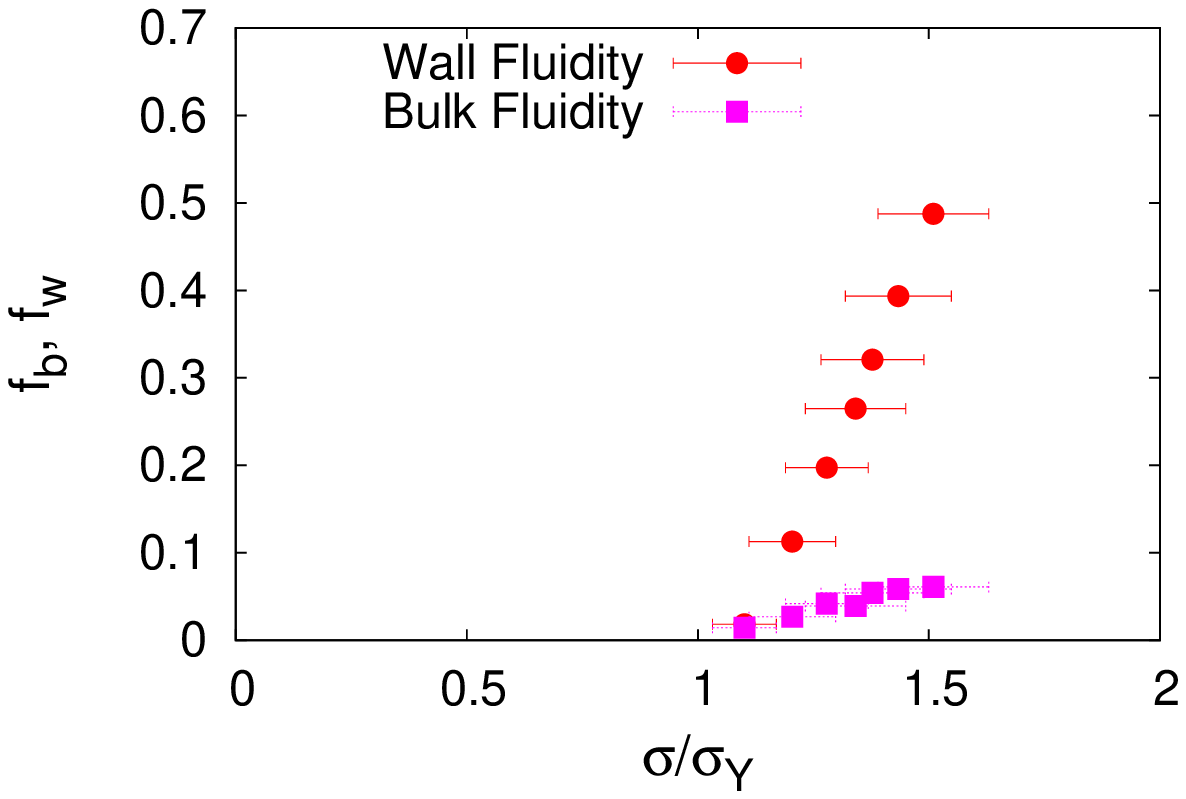}
\includegraphics[scale=0.42]{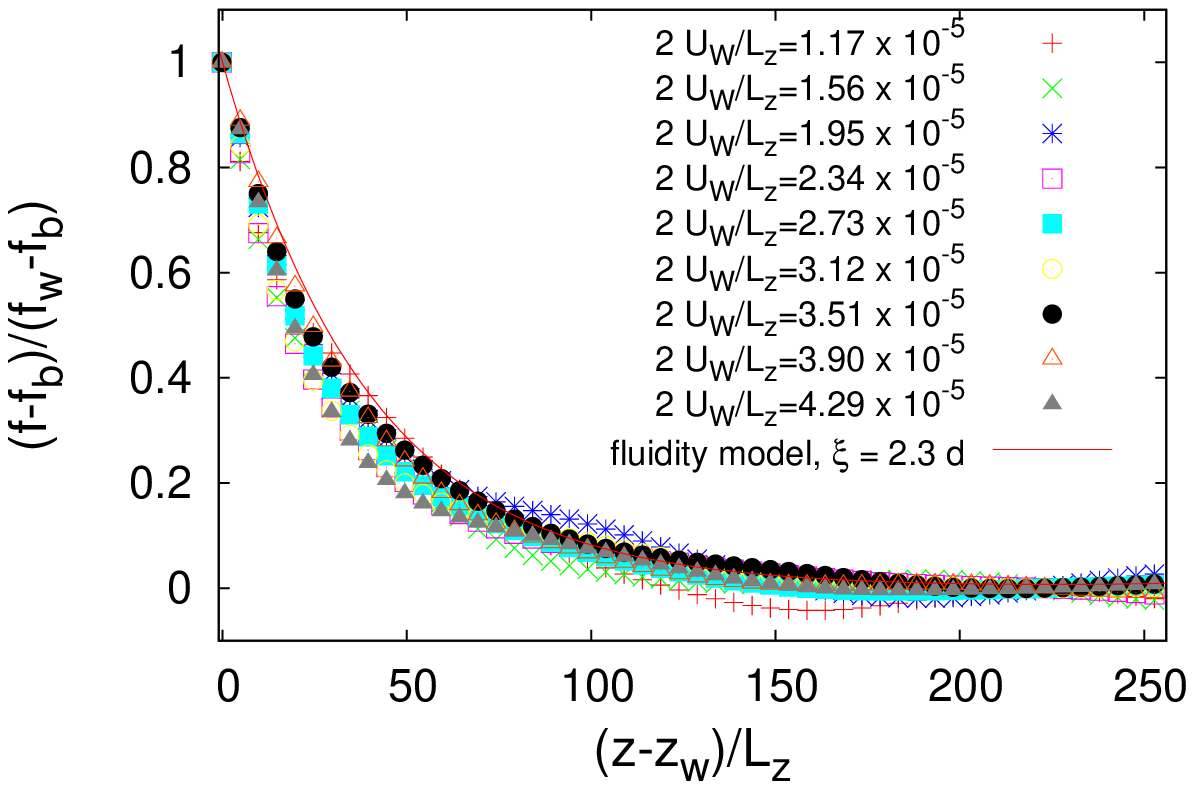}
\hfill
\caption{Left Panel: the average fluidity as a function of the distance from the walls in a Couette Flow simulation. Data are the same reported in the bottom panel of figure \ref{fig:1}. The vertical dotted line represents the distance from the wall at which we calculate $f_w$ in equation (\ref{eq:fluidityCF}). All the numerical simulations are performed above the yield stress $\sigma_Y$. Middle Panel: wall (w) and bulk (b) fluidity as a function of the normalized (with respect to the yield stress $\sigma_Y$) average stress in a Couette Flow simulation: bulk and wall rheology are different. Right panel: The fluidity shown in the left panel is reported and normalized with respect to the wall and bulk contributions, in order to extract the cooperativity scale according to equation (\ref{eq:fluidityCF}). \label{fig:6}}
\end{figure*}


Several remarks are in order.   First, the predictions for the velocity profiles exhibit features in qualitative agreement with the simulation results reported in the bottom panel of figure \ref{fig:1}: even though the shear stress is homogeneous, velocity profiles are not straight lines from wall to wall. Deviations from a linear profile close to the wall, extend over a characteristic distance fixed by $\xi$. The fluidity profiles for different values of the nominal shear rate $2U_W/L_z$ are reported in the left panel of figure \ref{fig:6}. All the numerical simulations are performed above the yield stress $\sigma_Y$. Starting from the wall region, the fluidity field decays towards the bulk value $f_{b}$, which can also be deduced from the rheological flow curve reported in figure \ref{fig:1}. As for the wall fluidity, $f_w$, we directly measure it at the distance evidenced by the vertical dots in the left panel of figure \ref{fig:6} and compare it with the bulk fluidity, $f_b$, in the middle panel of figure \ref{fig:6}. The existence of a specific wall rheology is clear: the wall fluidity is significantly larger than the bulk fluidity \cite{Goyon2}. To double-check the quantitative consistency with equation (\ref{eq:fluidityCF}), we rescaled all profiles with respect to the wall fluidity, $f_w$, and studied the quantity $(f(z)-f_{b}(\sigma))/(f_w-f_{b}(\sigma))$. The profiles of the rescaled fluidity collapse on the same curve, consistently with equation (\ref{eq:fluidityCF}) and a value of $\xi =2.3 \,d$. In line with the notion of cooperativity \cite{Goyon1, Goyon2}, describing the characteristic scale for non-local effects in the soft-glassy dynamics, we find that $\xi$ and the characteristic scale of plastic events, $\lambda_p$, are close to each other.

\section{Connection with Stress Correlation}\label{stresscorr}

\begin{figure}[h!]
\includegraphics[width=0.5\textwidth]{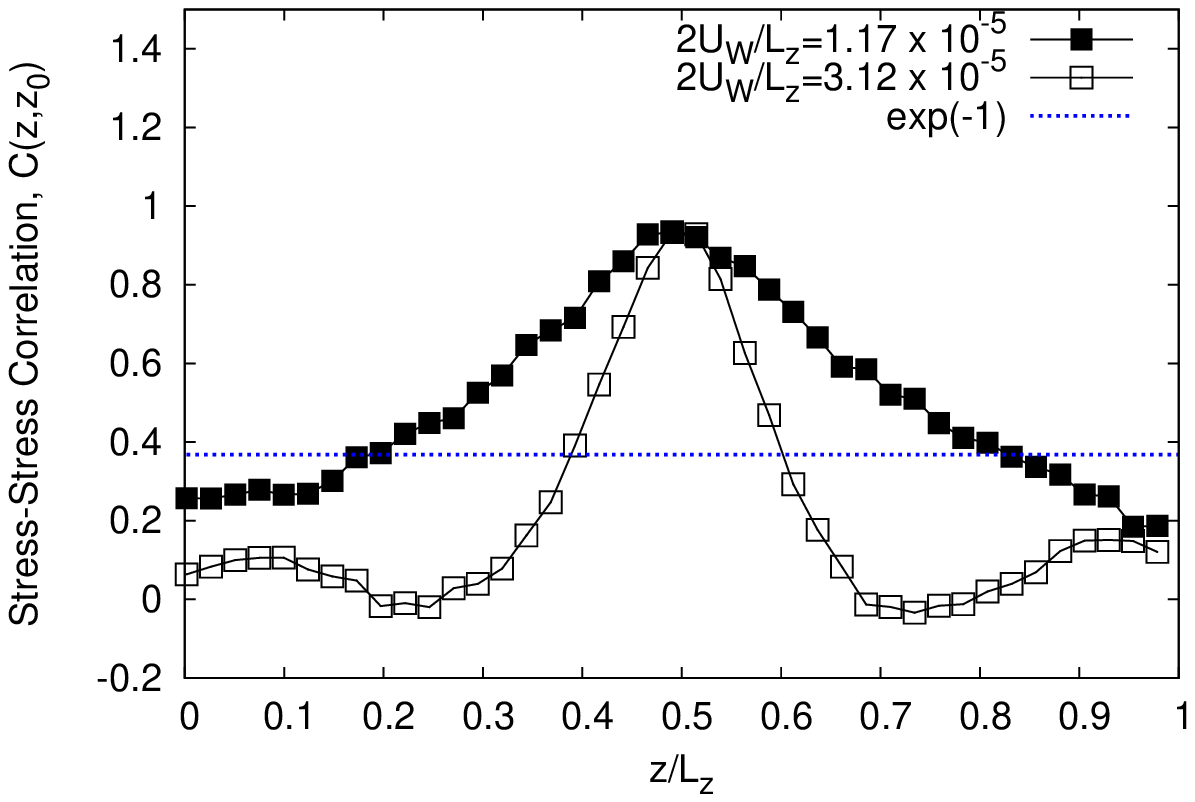}\\
\includegraphics[width=0.5\textwidth]{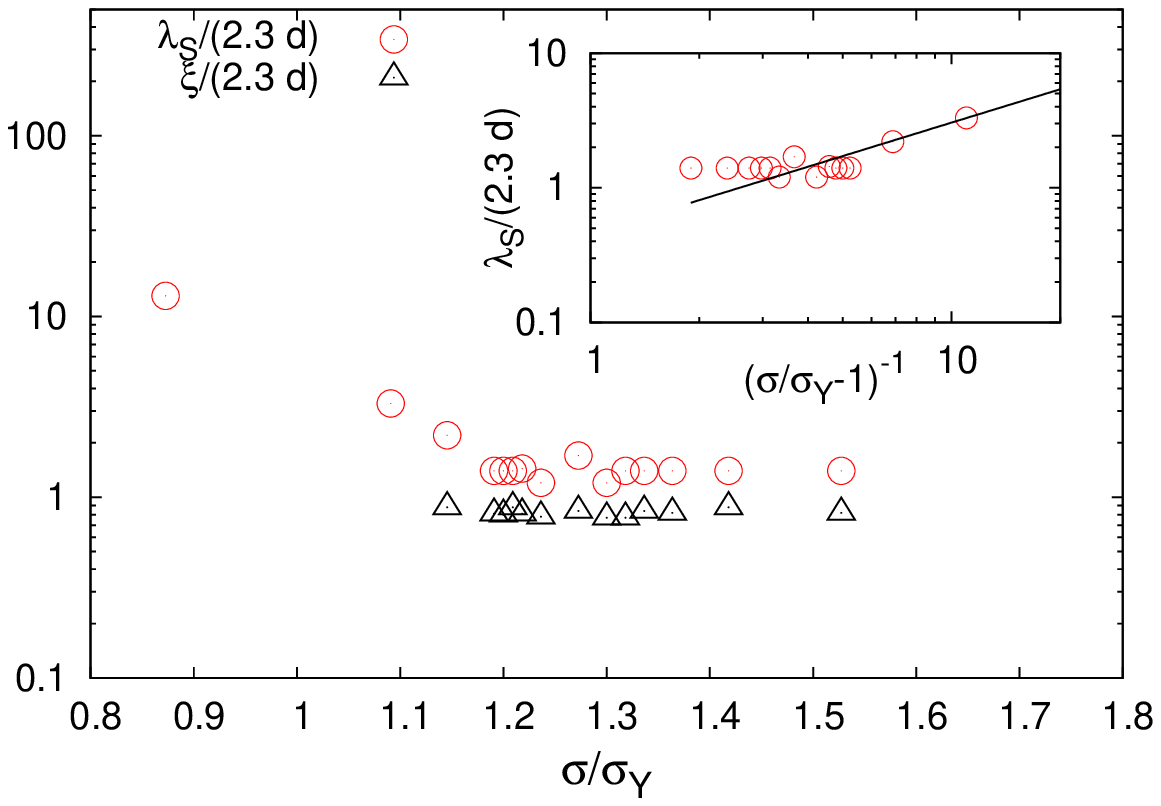}
\caption{Top panel: the stress correlation function $C(z,z_0)$, measured starting from $z_0=L_z/2$ (changing $z_0$ does not affect the conclusion). Open and Filled Symbols refer to different shear rates. Starting from the reference location (where, by definition, the stress correlation function is $C(z_0,z_0)=1$), we computed the stress correlation, $\lambda_S$, as the distance away from such location when the correlation function is $C(z_0+\lambda_S,z_0)=\mbox{exp}(-1)$. Bottom panel: the figure shows the cooperativity scale $\xi$ (triangles) as a function of the rescaled stress $\sigma/\sigma_Y$ in Couette Flow. The cooperativity scale $\xi$ is constant above the yield stress. In the same figure, we show the stress correlation scale, $\lambda_S$ (see top panel). The stress correlation scale shows an increasing trend at decreasing shears. Both the cooperativity and the stress correlation scales are normalized with $2.3 d$, where $d$ is the average droplet diameter. Inset:  we plot $\lambda_S/(2.3d)$ versus $\sigma_Y/(\sigma-\sigma_Y)$ in log-log scale; the solid line is the scaling prediction derived from the scalar model (\ref{m1}-\ref{m2}) (see text for details).  \label{fig:7}}
\end{figure}

We can gain further insights by studying the correlation of the stress in the material and exploring its connection with the results presented in sections \ref{plastic} and \ref{fluidity}. In particular, we measure the stress correlation scale $ \lambda_S$ in the system. Let $C(z,z_0)$ be the stress correlation function defined as:
\be
C(z,z_0) = \langle(\bar{\sigma}(z,t)\bar{\sigma}(z_0,t) \rangle_{t,c}
\ee
where $\bar{\sigma}(z,t)$ is the average of the stress along the mainstream direction, i.e. $\bar{\sigma}(z,t)= \langle \sigma(x,z;t) \rangle_x =\frac{1}{L_x} \sum_{x} \sigma(x,z;t)$, and where the subscript $c$ denotes the {\it connected} correlation function.  We estimate $\lambda_S$ as the distance away from the location $z_0=L_z/2$ where the correlation function is $C(z_0+\lambda_S,z_0)=\mbox{exp}(-1)$.  In the bottom panel of figure \ref{fig:7}, we plot $\lambda_S$. As one can appreciate, above the yield stress ($\sigma > \sigma_Y$), the stress correlation scale and the cooperativity scale $\xi$ are basically the same (up to a scale factor, close to $1$). However, very close to the yield stress, $\lambda_S$ shows a fast growth at decreasing shear. Such an increase of $\lambda_S$ can actually be explained by resorting to a very simple scalar rheological model for the stress field $\sigma(z,t)$ \cite{Picard,Takeshi,Janiaud06}:
\begin{eqnarray}
\label{m1}
\partial_t (\rho u) & =  \partial_z (\eta S  +  {\sigma})\\
\label{m2}
\partial_t {\sigma}  & = E S -  \frac{{\sigma}}{\tau}
\end{eqnarray}
where $u(z,t)=\langle u_x(x,z;t) \rangle_{x}$ is the average mainstream flow speed, $S=\partial_z u$ the shear, $\eta$ the molecular dynamic viscosity and $E$ the elastic modulus. Equation (\ref{m1}) is the momentum conservation relation, while equation (\ref{m2}) is a phenomenological model for the evolution of the stress.  Finally, $\tau$ is a relaxation time, diverging close to the yield stress \cite{recentEPL}. Such kind of models have been known for long in the literature \cite{Picard,Takeshi,Janiaud06}: equation (\ref{m1}) is usually considered in the stationary state, on account of inertia being totally negligible \cite{Picard}. In the stationary state, with an average stress $\sigma$ above the yield stress $\sigma_Y$, equation (\ref{m2}) is consistent with the Herschel-Bulkley global flow curve, equation (\ref{HB}) with
\begin{equation}
\label{tau}
\tau({\sigma}) = \frac{{\sigma}}{E S} = \frac{{\sigma} A^{1/\beta}}{E({\sigma}-\sigma_Y)^{1/\beta}}.
\end{equation}
Equations (\ref{m1}) and (\ref{m2}) can also be written as:
\begin{eqnarray}
\label{m1new}
\partial_t (\rho u) & = & \partial_z \Pi\\
\label{model}
\partial_t {\Pi}  & = &\left(E + \frac{\eta}{\tau}\right)S - \frac{\Pi}{\tau} + \frac{\eta}{\rho} \partial_{zz} \Pi
\end{eqnarray}
where $\Pi = \eta S + {\sigma}$.  Finally, ignoring the inertial term represented by (\ref{m1new}), we can refer to  (\ref{model}) in order to understand the behavior of the stress correlation functions. Let us remark that, in most cases, such as those presented here, the molecular viscosity $\eta$ is much smaller than the ``solid'' contribution and we can estimate $\eta_{eff} \sim \sigma/S = E \tau \gg \eta$. This ensures a negligible difference between $\Pi$ and $\sigma$. Equation (\ref{model}) shows that the stress correlation scale should be of the order $\sqrt{\eta \tau/\rho}$, which diverges close to $\sigma_Y$ in agreement with our findings.  In particular, by using equation (\ref{HB}), we can predict $ \lambda_S \sim \sqrt{\eta \tau/\rho} \sim 1/(\sigma-\sigma_Y)^{1 \over {2\beta}}$. In the inset of the bottom panel figure \ref{fig:7}, we plot $\lambda_S/(2.3d)$ versus $\sigma_Y/(\sigma-\sigma_Y)$ in log-log scale, which shows that our prediction is consistent with numerical data with ${1 \over {2 \beta}} \sim 0.82$.  Close to the yield stress, we cannot measure $\xi$ by using equation (\ref{eq:fluidityCF}), since the fluidity near the wall becomes close to the bulk fluidity and both tend to zero. We were able to obtain accurate measurements only down to $\sigma/\sigma_Y \approx 1.1$, where the cooperative length $\xi$ does not show any substantial variation (see triangles in the bottom panel of figure \ref{fig:7}). The computation of $\lambda_S$, instead, does not result from any best fit procedure and it is an {\it independent} measure of space correlations.  We note that the increase of $\lambda_S$ near $\sigma_Y$ is also consistent with the results shown in figure \ref{fig:5}, suggesting that below the yield stress, the system behaves as an elastic medium with long-range order, where plastic events occur without any preferential location.


\begin{figure}[h!]
\includegraphics[scale=0.53]{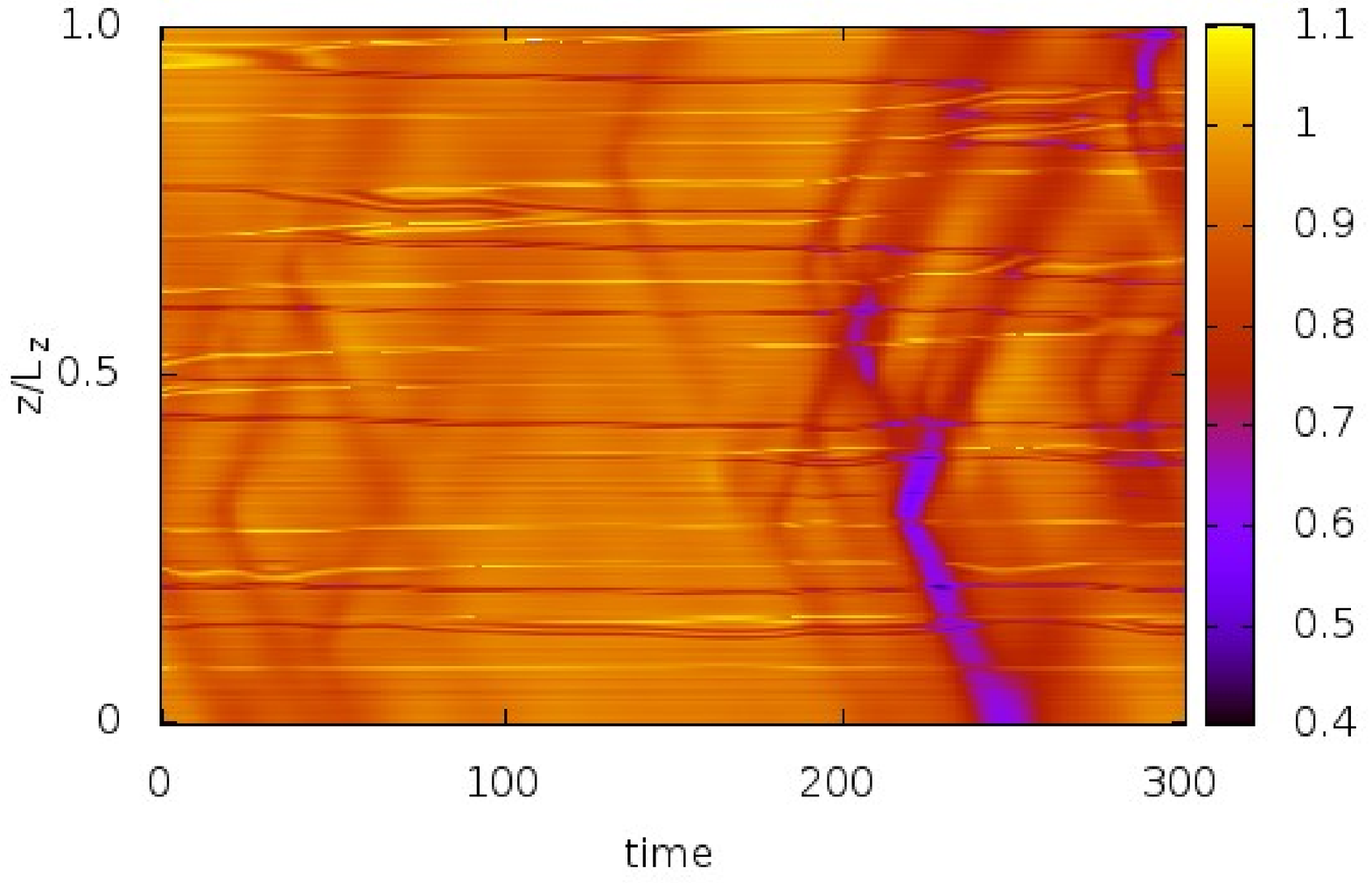}
\includegraphics[scale=0.7]{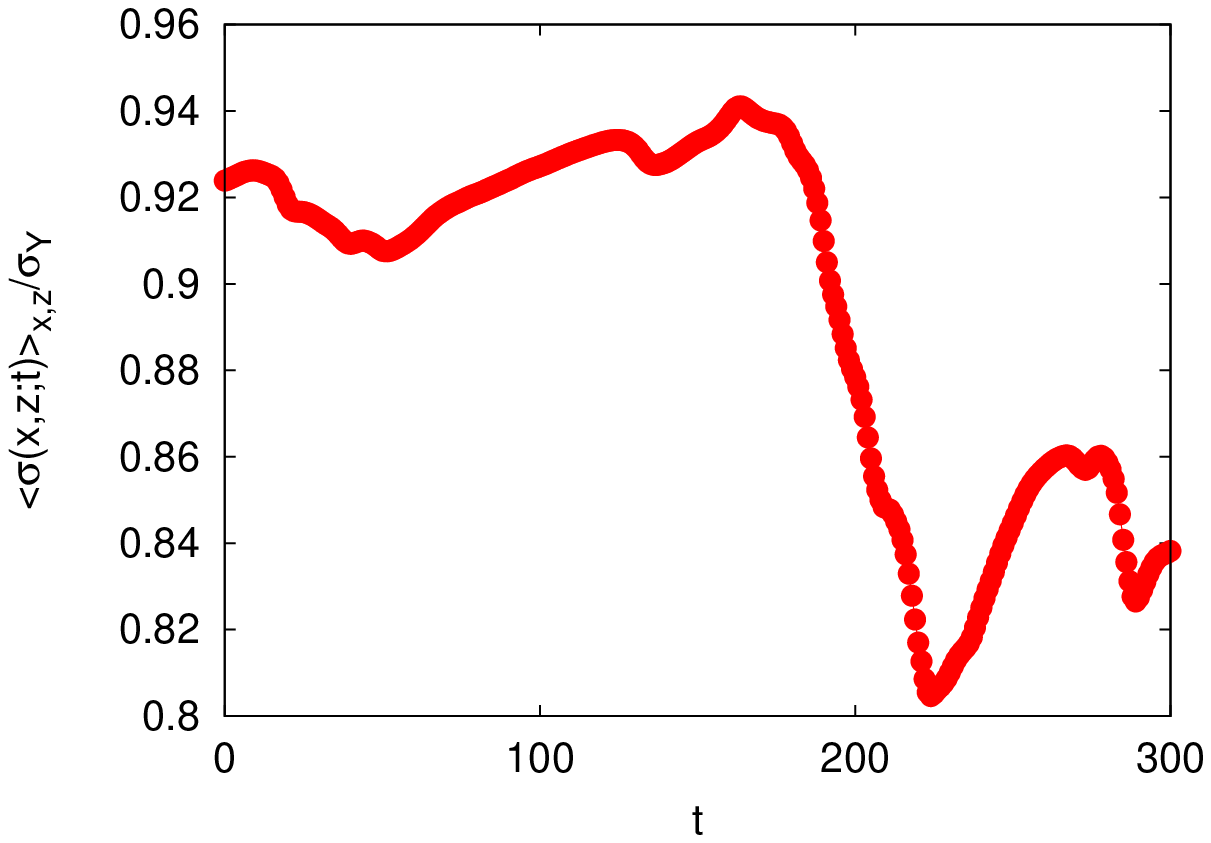}
\caption{The figure highlights the time/space dynamics of the stress in a Couette Flow for $\sigma/\sigma_Y=0.88$. Top Panel: we consider the stream-flow averaged stress $\bar \sigma(z,t) = \langle \sigma(x,z;t) \rangle_x =\frac{1}{L_x} \sum_{x} \sigma(x,z;t)$ in a time interval $[0:300]$ (in units of $1000$ lbu). The vertical axis is the wall-to-wall distance $z$ and the horizontal axis is time $t$. The stress is normalized with the yield stress $\sigma_Y$ (see figure \ref{fig:1}). Bottom panel: we report the $z$-averaged of $\bar{\sigma}(z,t)$ (again normalized with $\sigma_Y$), as a function of $t$, in the same time interval of the top panel. The interesting point is the neat evidence of propagation of elastic waves associated to plastic events. This phenomenon is clearly detectable in correspondence with the sudden drop in the global stress at $t \sim 200$. \label{fig:2}}
\end{figure}

\begin{figure}[h!]
\includegraphics[scale=0.712]{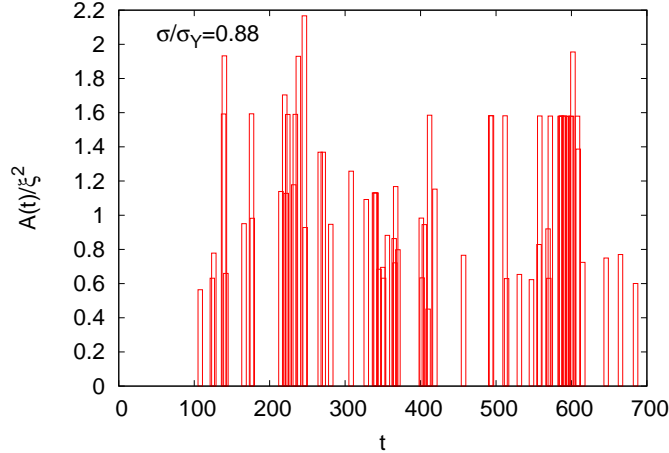}
\caption{Time dynamics of plastic events in Couette Flow simulation at $\sigma/\sigma_Y=0.88$. We report the area of the plastic events $A(t)$ rescaled by $\xi^2$ in a time interval $[0:700]$ (in units of $1000$ lbu). The interesting point is the neat clustering of the plastic events.  \label{fig:8}}
\end{figure}

\begin{figure}[h!]
\includegraphics[width=0.6\textwidth]{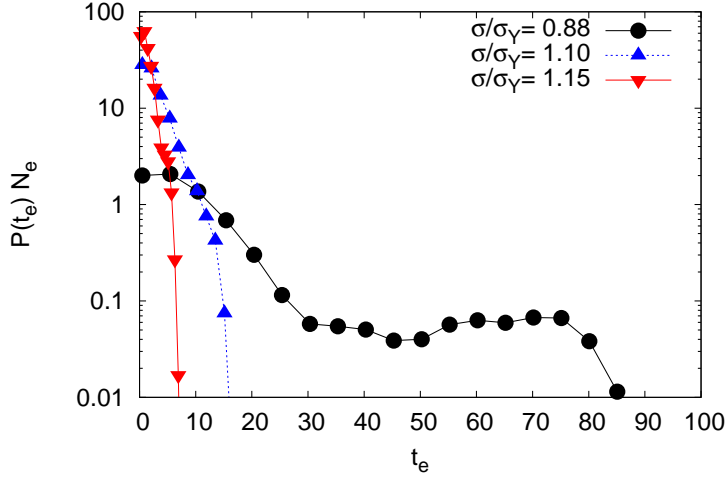}
\caption{Probability density distribution $P(t_e)$ of the time $t_e$ between two consecutive plastic events in a Couette Flow simulation for $\sigma/\sigma_Y = 0.88,\, 1.1,\, 1.15$ (log-linear scale). The values of $P(t_e)$ are multiplied by the number of events $N_e$ observed during the period of $10^6$ time steps, where $N_e= 89$, $N_e=288$ and $N_e=600$ respectively. The striking feature is the long tail of $P(t_e)$ observed for $\sigma/\sigma_Y =0.88$ which shows the time intermittent dynamics of the plastic events. \label{fig:9}}
\end{figure}

Other key signatures of the physics below the yield stress are provided in figures \ref{fig:2} and \ref{fig:8}.  In figure \ref{fig:2}, we monitor the space-time distribution of the stress-field, $\sigma(x,z;t)$, as well as its average along the mainstream direction, $\bar \sigma(z,t)$, in a Couette Flow for $\sigma/\sigma_Y=0.88$. The diagonal stripes in the top panel of figure \ref{fig:2} provide a neat signature of propagating stress-waves, which become apparent in close connection with the drop of the average stress, as shown in the bottom panel. This shows that the dynamics of the system supports propagation of stress-waves, in connection with the occurrence of stress-releasing plastic events. Plastic events also show an intermittent clustering in time, as evidenced in figure \ref{fig:8}, where we report the area $A(t)$ related to plastic events in a Couette Flow simulation at $\sigma/\sigma_Y=0.88$. As we can see, a substantial number of plastic events occur in quite short time intervals, after which quiescent periods are observed. It is tempting to speculate that there are ``avalanches'' of plastic events. It seems that the stress-waves generated by the first plastic event, trigger a number of other events, each generating stress-waves and,  eventually, triggering further plastic events \cite{Picardb}. Such intermittent clustering in time can be indeed quantified by looking at the probability density distribution $P(t_e)$, $t_e$ being the time interval between two successive plastic events. In figure \ref{fig:9}, we show $P(t_e)$ for $\sigma/\sigma_Y = 0.88$, $\sigma/\sigma_Y =1.1$ and $\sigma/\sigma_Y=1.15$. A striking feature emerges from figure \ref{fig:9}: the clustering properties of plastic events are peculiar of the pre-yield condition. Only for $\sigma < \sigma_Y$, we observe a long tail in $P(t_e)$ which is a clear signature of time intermittency or clustering in the plastic events. Actually, we observe that the tail increases in the course of the numerical simulations; i.e. the system shows aging.


\section{Interface Correlations and Elastic Stress}\label{ELASTIC}

Although the simple model reported in equation (\ref{model}) seems to explain the behavior of $\lambda_S$ near the yield stress, it does not reveal much of the underlying physics. Basically, the statement $\lambda_S \rightarrow \infty $ or $\tau \rightarrow \infty $ is a shorthand to characterize the yield stress transition. A more interesting question concerns the physical mechanism characterizing the transition, i.e. the reason why the stress correlation scale increases and/or the relaxation time increases. In this section, we provide a quantitative answer to this question by further exploring the space-time correlations of the elastic stress of the system. We concentrate on the space-time correlations of the motion of the interface which, we argue, are responsible  for  the increase in the stress correlation scale discussed in the previous section. As shortly outlined earlier on, the picture we have in mind is the following: very close to the yield stress, plastic events take place in an otherwise elastic material \cite{Picardb,Picardbb}. During a plastic event, the whole interface moves and changes the local stress fluctuations, as well as the interface configuration. Because of the effect of the stress waves, which propagate {\it after } the end of the plastic event, the interface may become locally unstable and there is a relatively high probability to trigger further plastic events \cite{krysac}. Since the motion of the interface induces a large change in the stress fluctuations, the stress correlation scale is large (order the system size). Our qualitative description highlights the link between an increase  in the relaxation time of the system ($\tau$) and the increase of space correlation.


\begin{figure*}[t]
\includegraphics[width=0.45\textwidth]{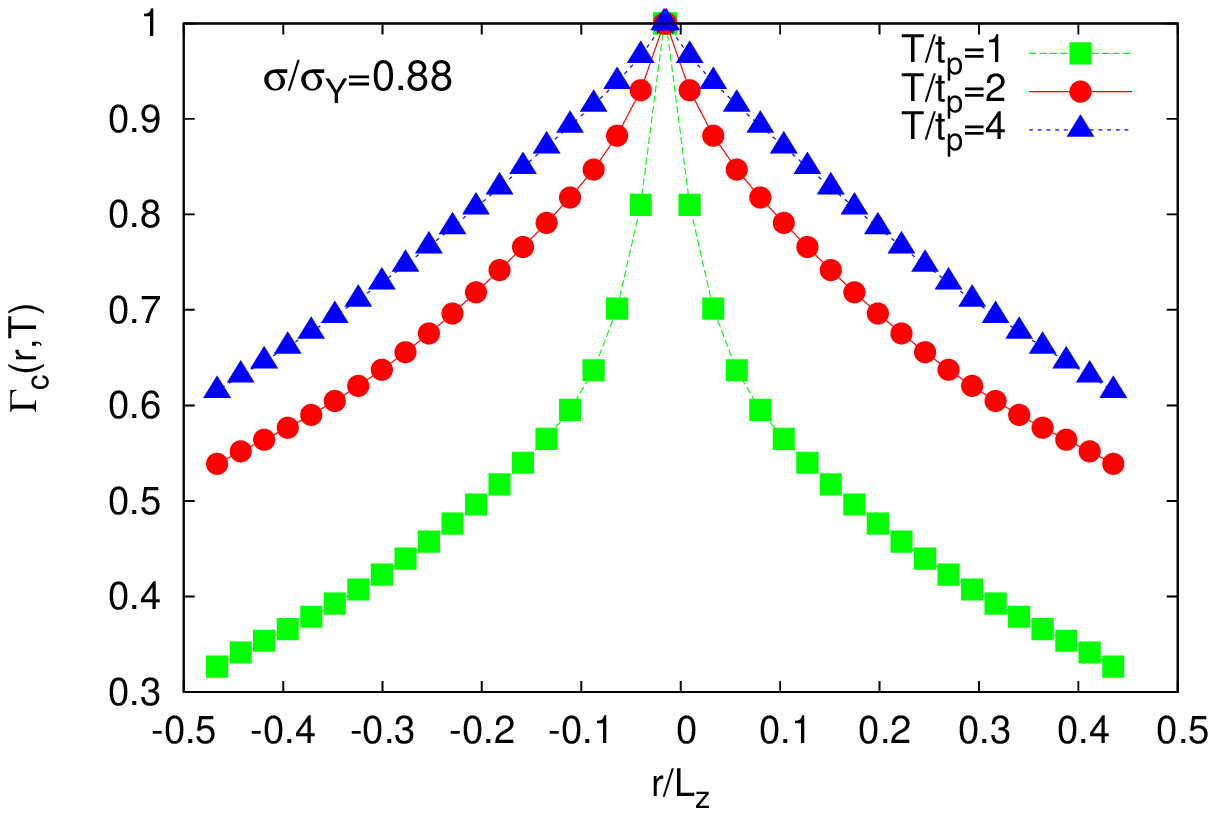}
\includegraphics[width=0.45\textwidth]{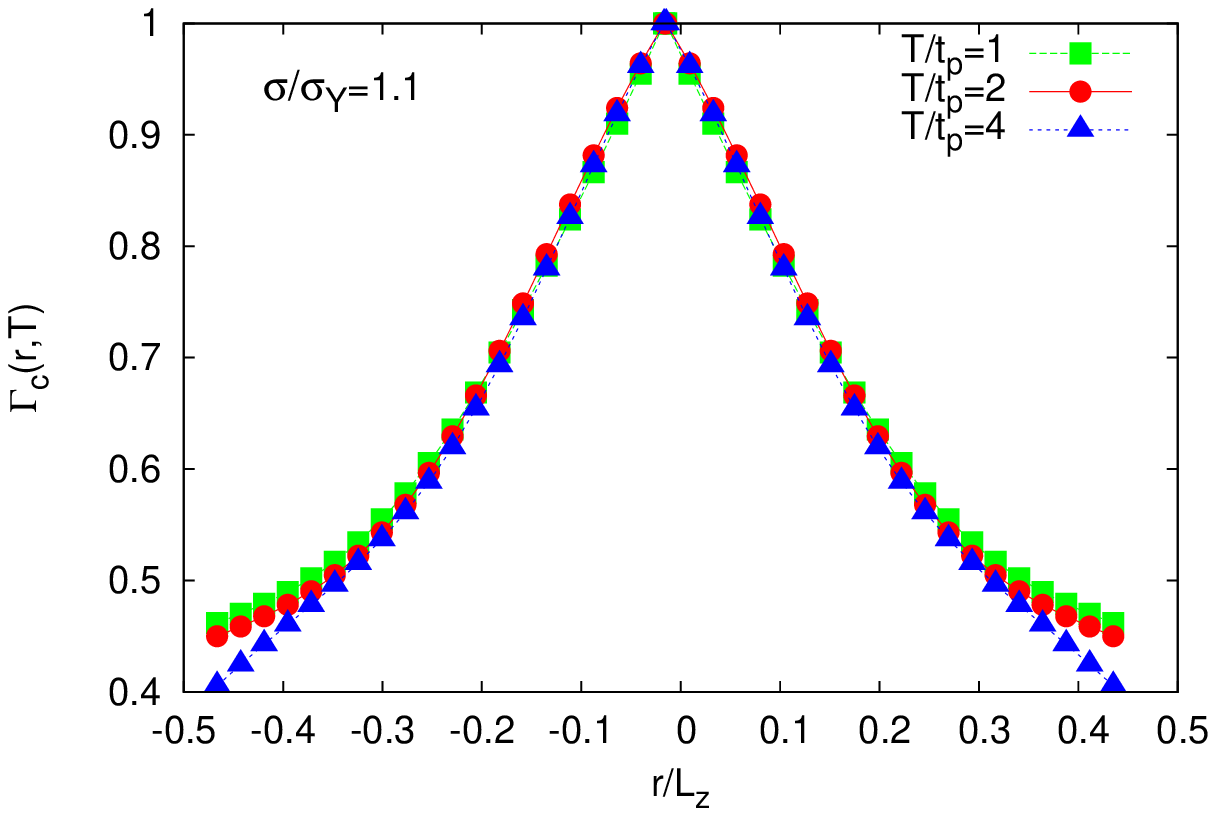}
\caption{The figure shows $\Gamma_c(r,T)$, i.e. the connected correlation function of the overlap (see text for details), for $\sigma/\sigma_Y=0.88$ (left panel) and $\sigma/\sigma_Y=1.1$ (right panel) with $T/t_p=1,2,4$, $t_p$ being the characteristic time of the plastic event. Given two field configurations separated by a time $T$, $\Gamma_c(r,T)$ provides a measure of the spatial correlation existing between two points separated by the distance $r$. When $\sigma/\sigma_Y=0.88$, we observe a large spatial correlation which increases in time due to propagation of stress-waves. \label{fig10}}.
\end{figure*}

To define a quantitative measure of the correlations associated with the motion of the interface,  we introduce the phase field $\phi(x,z;t) \equiv \rho_A(x,z;t)-\rho_B(x,z;t) - \langle(\rho_A-\rho_B) \rangle_{x,z}$. Next, we define the {\it overlap} ${q}(x,z;t,t+T)$ as:
\begin{equation}\label{overlap0}
{q}(x,z;t,t+T)  =  \frac{\phi(x,z;t)\phi(x,z;t+T)}{\bar{\phi}(t)\bar{\phi}(t+T)}
\end{equation}
where $\bar{\phi}^2(t) \equiv \langle \phi^2(x,z;t) \rangle_{x,z}$. The physical meaning of ${q}$ is the following: for constant $T$, let us indicate by ${q}_{x, z, t} (T) = \langle {q} (x,z;t,t+T) \rangle_{x, z, t}$ the space-time average of ${q}$; then ${q}_{x, z, t} (T)$ provides a quantitative measure of how much two field configurations, separated by a time $T$, are on average correlated. Thus, to compute space-time correlations, we need to evaluate the space correlation of ${q}$
\begin{equation} \label{gamma}
\begin{split}
\Gamma(r,T) =  \langle & {q}(x,z+r;t,t+T) {q}(x,z;t,t+T)+ \nonumber \\
& {q}(x+r,z;t,t+T) {q}(x,z;t,t+T) \rangle_{x,z,t}
\end{split}
\end{equation}
where $-L_z/2 \le r \le L_z/2$. Since the change in the configuration of the phase field is due to the interface motion, $\Gamma(r,T)$ is a quantitative measure of the space-time correlations of the interface dynamics. By using the Voronoi construction, we have identified plastic events as changes in the topological configuration of the interface. Such changes are actually {\it instantaneous}. However, a careful inspection of the dynamics shows that the interface motion associated to a local plastic event, takes a finite time $t_p$. The value of $t_p$ is not fixed, although it does not show large variations among different plastic events.  The characteristic time $t_p$ can be estimated of the order of $\lambda_p/v$, where $v$ is the stress-wave velocity: for a time scale much longer than $\lambda_p/v$ it is unlikely that any locally confined source of energy does not radiate out the region where the plastic event occurs. A few numbers may help elucidating the picture. Using $\lambda_p = 100$ and $v=0.02$ in lbu, we obtain $t_p \approx 5000$ lbu. Note that the time for a stress-wave to propagate from one boundary to the other is $t_E \sim 10 \, t_p$, whereas the time scale induced by the external driving is in the range $[30:100]\,t_p$ across the yield  stress transition, where the longer time refers to the value at  $\sigma/\sigma_Y = 0.88$. We then consider $\Gamma_c(r,T)$ (the connected correlation function of $\Gamma(r,T)$) for $\sigma/\sigma_Y = 0.88$ (figure \ref{fig10}, left panel) and $\sigma/\sigma_Y=1.1$ (figure \ref{fig10}, right panel) and for $T/t_p=1,2,4$. For $T \sim t_p$, based on the qualitative picture previously described, a clear correlation is expected. Note the long tail in the correlation function for large $r$ at $T/t_p=1$: this quantitative measure indicates that the whole interface is spatially correlated on time scales smaller than $L_z/v$ and comparable to the time scale of the plastic event $t_p$. However, for $\sigma/\sigma_Y =0.88$, the correlation increases with time due to the propagation of stress waves, whereas for $\sigma/\sigma_Y=1.1$ the long tail in the correlation length disappears. Moreover, when the system starts to flow at $\sigma/\sigma_Y=1.1$, stress waves no longer propagate and consequently one cannot observe long-range correlations in the interface motions. Figure \ref{fig10}, therefore, supports our view and indicates the interface motion as the source of the large scale correlation in the stress fluctuations.


\section{Summary and outlook}

We have presented quantitative measurements of the statistics and correlations of plastic events, as they arise in the proximity of the yield-stress threshold, obtained by using simulations of concentrated emulsion droplets under soft-glassy conditions. We provide two basic results. {\it First}, above the yield stress, the typical spatial scale of the plastic events, $\lambda_p$, is in a good quantitative match with the cooperativity scale, $\xi$, introduced by previous authors \cite{Goyon1,Goyon2}. Both scales are close to the correlation scale of the fluctuating stress within the material, $\lambda_S$. Among others, a notable result emerging from the above  findings is the spontaneous segregation of the plastic events within  a near-wall layer of thickness $2 \lambda_p$. {\it Second}, below the yield stress, $\lambda_S$ shows a clear increase and plastic events exhibit intermittent clustering in time, while showing no preferential locations. This is understood in terms of the long-range {\it amorphous} order emerging at the yield stress threshold, where one cannot purport the system as an assembly of mesoscopic elements: the whole interface configuration comes into play during plastic events and the ``energy landscape'' should be classified in terms of interface configurations with large space-time correlations.\\
Another important aspect emerging from our analysis is the key role of stress-waves.  Usually, having slow flows of soft-glassy materials in mind, one neglects  inertial effects in developing mesoscopic models for elasto-plastic materials \cite{Picard,Picardb}. In this work, inertia is {\it not} invoked to explain the non-linear rheology of the  system, but to allow the propagation of sound waves in the solid, which proves key to sustain long-range dynamic correlations. At low shear rates, experiments are performed to ensure a uniform strain in the system and a nearly constant stress. This is certainly the case when one considers linear rheology in a Couette Flow configuration at very low frequency. Also, the computations performed with the Oscillatory Strain display a clear uniform rate strain and uniform stress for small $\sigma_P$. However, close to the yield stress, space   fluctuations of the stress and the interfaces are crucial to   correctly describe the dynamics of the system. As we have seen,   stress-waves are able to trigger plastic events and produce an  avalanche. Stress-waves can exist only by assuming the active presence of inertial terms. As a matter of fact, mesoscopic models which  describe the deformation of elastic solids, do make use of inertia terms \cite{Dahmen09,Takeshi}. A recent study by Salerno \& Robbins\cite{Robbins} shows indeed that inertia can strongly influence activity bursts and avalanches in sheared disordered solids.\\ 
Overall, all the simulation results presented in this paper refer to a situation where $t_D>t_E>t_c$, with $t_D=L_z^2\rho/\eta$ the diffusive time associated with molecular viscosity (see equations (\ref{m1})-(\ref{m2})), $t_E$ the elastic time for a stress-wave to propagate from one boundary to the other (see section \ref{ELASTIC}) and $t_c \approx \omega_c^{-1}$, where $\omega_c$ is the frequency  at which the storage modulus $G^{\prime}(\omega)$ and the loss modulus $G^{\prime \prime}(\omega)$ cross each other, i.e. $G^{\prime}(\omega_c) \approx G^{\prime \prime}(\omega_c)$.  In our case, $t_D/t_E \approx 10$ and $t_c$ is found to be of the order of the characteristic time of plastic events $t_p$ (see section \ref{ELASTIC}), with $t_E/t_p \approx 10$. A close look at some experimental data\cite{Goyon1,Goyon2,mason2}, reveals that $t_D/t_E$ is in the range $[2:20]$ and $t_E/t_c$ in the range $[1:10]$, thus suggesting that the adopted ordering of time scales is reasonable.\\ 
Finally, we wish to highlight the importance of ``randomness'' and disorder in the initial condition \cite{Katgert13}, which provides a nontrivial feedback to the dynamics. All the simulations presented here have been performed with a small but not negligible polydispersity in the initial configuration. For an ordered hexagonal packing of monodisperse droplets, the yield stress and strain follow from Princen theory \cite{Princen,Kraynik}. Even a small polidispersity changes the yield strain and opens the way to a much richer and complex dynamics. However, the role of polydispersity or space randomness in the system is still not clearly understood.  In particular, preliminary results suggest that an increase in the polidispersity is equivalent to increase the level of ``noise'' in the system and change the space-time correlations.  Although most of the above discussions are rather speculative, we argue that our work may enhance the interest in discussing space-time correlation near the yield stress transition and provide some insights to develop a complete theory of soft-glassy rheology. \\ 
We are particularly grateful to A. Scagliarini for his careful reading of the manuscript. M. Sbragaglia \& R. Benzi kindly acknowledges funding from the European Research Council under the Europeans Community’s Seventh Framework Programme (FP7/2007-2013)/ERC Grant Agreement No. 279004. P. Perlekar \& F. Toschi acknowledge partial support from the Foundation for Fundamental Research on Matter (FOM), which is part of the Netherlands Organisation for Scientific Research (NWO).

\bibliography{rsc}       
\bibliographystyle{rsc} 
\end{document}